\documentclass[11pt, oneside]{article}
\usepackage{geometry}
\usepackage[parfill]{parskip}
\usepackage{hyperref}
\usepackage{authblk}

\usepackage{paralist} 
\usepackage[hang,flushmargin]{footmisc}
\usepackage{natbib}

\title{Health and Demographic Surveillance Systems \\and the 2030 Agenda: Sustainable Development Goals}
\date{\today}  

\author[1,2,*]{Samuel J. Clark}

\affil[1]{Department of Sociology, The Ohio State University}
\affil[2]{MRC/Wits Rural Public Health and Health Transitions Research Unit (Agincourt), \protect\\ School of Public Health, Faculty of Health Sciences, University of the Witwatersrand}
\affil[*] {Contact: work@samclark.net, +1 (206) 303-9620}

\begin{document}
\maketitle

\begin{abstract}

\noindent The health and demographic surveillance system (HDSS) is an old method for intensively monitoring a population to assess the effects of healthcare or other population-level interventions - often clinical trials.  The strengths of HDSS include very detailed descriptions of whole populations with frequent updates.  This often provides long time series of accurate population and health indicators for the HDSS study population.  The primary weakness of HDSS is that the data describe \textit{only} the HDSS study population and cannot be generalized beyond that.

\vspace{5pt}

\noindent The 2030 agenda is the ecosystem of activities - many including population-level monitoring - that relate to the United Nations (UN) Sustainable Development Goals (SDG). With respect to the 2030 agenda, HDSS can contribute by: 
\begin{inparaenum}[(1)]
\item continuing to conduct cause-and-effect studies; 
\item contributing to data triangulation or amalgamation initiatives;
\item characterizing the bias in and calibrating \textit{big data}; and 
contributing more to the rapid training of data-oriented professionals, especially in the population and health fields.
\end{inparaenum}
\end{abstract}

\section{Preamble}

This note is an edited version of a paper and presentation by the author delivered at a United Nations Population Division experts' group meeting titled `Strengthening the Demographic Evidence Base for the Post-2015 Development Agenda' that took place in New York, USA October 5-6, 2015.

\section{Inroduction}

Over the past decades health and demographic surveillance systems (HDSS) have proliferated in low- and middle-income countries (LMIC) as population-based research platforms designed to develop and test public health-oriented interventions.  Many survive their founding project and grow into general purpose population and health research institutions that contribute new evidence and findings in a wide variety of fields.

This note briefly describes the HDSS method with a summary of its strengths and weaknesses and then speculates on how HDSS might work together with other traditional and new sources of data and analytical approaches to strengthen the demographic, health and development evidence base in the context of the United Nations (UN) Sustainable Development Goals (SDG) whose targets should be met by 2030. 

\section{Health and Demographic Surveillance Systems - HDSS}

\subsection{HDSS as a Method}

The fundamental motivation for HDSS is the need to accurately account for the full population \textit{at risk} of something (e.g. dying, being infected with a specific disease, etc.) or participating in a trial of some kind.  With that in mind, a population is clearly defined and regularly monitored and all the in/out flows of people to/from the population are fully documented.  The population is typically defined as \textit{everyone} living within a geographical region, and the flows are \textit{birth} and \textit{in-migration} for the `ins' and \textit{death} and \textit{out-migration} for the `outs'.  

After an initial census, each household within the demographic surveillance area (DSA) is visited at least annually and a detailed set of information on each household member is updated.  Most HDSS sites conduct so-called `update' rounds on either a quarterly, 3-monthly or semi-annual basis.  This provides frequent updates on individual people, the households they live in and their community as a whole.  The obvious balance that must be struck is between round frequency and cost, and the key factors that drive that balance are how well the site wants to characterize pregnancy-related outcomes and early child death, especially miscarriages, still births and neonatal mortality.  To do this well, the round frequency needs to be short enough to identify all pregnancies and their outcomes.

This basic design creates an observational platform capable of extremely intensive monitoring with respect to time, space and a wide variety of social/medical dimensions.  HDSS data are \textit{prospective, densely linked and very detailed}.  This provides the opportunity for simultaneous disaggregation along many dimensions, and more unusual and useful, potential study of \textit{cause-and-effect} because the same entities (people, households, communities) are followed through time.

Although the details vary widely from site to site, HDSS typically includes data on basic demography\footnote{Vital rates, migration, sex-age structure.}, socio-economic status (SES) through household asset information\footnote{In most cases adapted from World Bank/DHS household asset surveys.}, cause of death (COD) through verbal autopsy (VA)\footnote{Often using either the INDEPTH Network standard VA instrument or the WHO standard VA instrument.}, various biomarkers and a wide variety of other social and biomedical data.

The typical population size for an HDSS site is about 80,000 individuals in about 12,000 households.  Long-running HDSS sites, such as Matlab in Bangladesh and several in West Africa, have been operating since the 1960--70s, and there are a number in East and Southern Africa that are  more than 20 years old.  

\subsection{Strengths and Weaknesses of HDSS}

HDSS data are likely the most timely and detailed of all population/health data regularly generated in LMICs, and the HDSS platform is often used to conduct rigorous randomized, controlled trials (RCT) and less rigorous observational studies of cause-and-effect.  The accumulated population and health data generated by HDSS sites are often used to conduct detailed retrospective population-based studies that include both sexes, all ages, all SES levels -- i.e. full disaggregation.  Data like these are usually extremely rare in LMICs\footnote{For example, prospectively collected data on all-age mortality.}.

This very level of intensity and detail is also the main challenge associated with HDSS.  The HDSS study design is a nominally 100\% census of a geographically-defined population and therefore does not \textit{represent} a any larger population in the sense of a sample survey.  Consequently results produced from HDSS data \textit{cannot} be generalized to larger populations, although it is very tempting to do so.  This lack of a statistical design that guarantees generalizability is one of two very significant challenges for HDSS data. 

The second is the \textit{Hawthorne Effect}\footnote{The `observer' effect in which the act of observing people changes their behavior above and beyond active intervention.}.  HDSS study communities are observed comprehensively for long periods of time and participate in multiple, often overlapping, trials explicitly aimed at changing their health or behavior.  Even if adequate control groups are maintained for \textit{a} study, those people are involved in other studies, and over the course of decades of being studied, it is certain that there is no unaffected control group, even if that were feasible from an ethics perspective.

The big strengths of HDSS data are their temporal coverage, detail, dense linkage and the fact that they exist at all for the chronically under-documented populations in LMICs where HDSS sites operate.  

The important weaknesses of HDSS data are their lack of generalizability, brought about by the lack of a statistical framework that relates HDSS study populations to a larger population and by the Hawthorne Effect, the fact that they are intensely observed, and this in and of itself, modifies the population.

%
%
%
%
%
%
%
%
%
%

\subsection{Networks of HDSS Sites}

There are two prominent networks that coordinate some of the activities of HDSS sites.  The INDEPTH Network is an NGO based in Accra, Ghana that provides a clearinghouse function for HDSS activities, coordinates funding and activities of some large multi-site projects, conducts regular scientific meetings for its member sites and provides a wide range of resources to HDSS sites.  The INDEPTH web site \url{http://www.indepth-network.org} provides information on the network and its members sites and is updated continuously.  

The ALPHA Network is a group of HIV surveillance sites in East and Southern Africa that includes a number of HDSS sites that collect HIV biomarkers.  The ALPHA Network is organized by members of the Department of Population Health at the London School of Hygiene and Tropical Medicine in London and maintains a strict focus on HIV-related health and population investigations.  The ALPHA Network maintains standardized data sets contributed by each member site and conducts analytical/training workshops to produce publications on HIV-related questions.  The ALPHA Network web site \url{http://alpha.lshtm.ac.uk} is updated frequently and provides additional information.

\subsection{Insights from HDSS}

Over decades HDSS sites have contributed significant insights in many areas ranging from basic demographic indicators to specific trial results related to vaccines, dietary supplements, mosquito bed nets, contraceptive effectiveness, HIV treatment and prevention effectiveness, effects of different approaches to health care system design, and many more.  Because the data are so rich, they can be used in very many new combinations and consequently speak to a very wide range of health and social questions, and often, the intersection of those questions - such as the effects of different approaches to health care systems.  The INDEPTH Network web site has a publications page that lists many recent published articles, but there are many more that can be found with more general searches using the names of individual HDSS sites. 

%
%
%
%

\subsection{Availability of HDSS Data}

Because they are densely-linked\footnote{Through time, across space and among different individuals and groups of individuals.}, prospective data updated on a frequent basis, HDSS data are comparatively complex.  This requires HDSS-specific data models, often implemented using relational database management systems.  Consequently, HDSS data are difficult to present in an easily understandable form, and it is difficult or not possible to adequately anonymize HDSS data for open access.  These factors have prevented or slowed down the movement to public access that is becoming more common with other types of population-based data.  

There are three ways to access HDSS data: 
\begin{inparaenum}[(1)]
\item by contacting the site directly and negotiating access to data; 
\item by working through either of the two prominent networks of HDSS sites -- the INDEPTH Network or the ALPHA Network, above; and
\item by accessing HDSS data on the INDEPTH Network's iShare data repository \url{http://www.indepth-ishare.org}. iShare contains a growing set of anonymized, individual-level data from HDSS sites in a publicly available form.
\end{inparaenum} 

For most serious studies, data must be requested directly from the HDSS sites.  This process requires understanding what data are available and their general structure and then making a request to an expert data manager at the site who writes a customized structured query language (SQL) script to extract and manipulate the raw data into the form necessary for the study.  This is an \textit{ad hoc}, time consuming process that makes taking full advantage of HDSS data difficult. The INDEPTH Network data repository is working to make this process less tedious and much quicker. 

Finally, there are thousands of published articles using HDSS data that provide additional results, including a growing series of `cohort profile' articles published in the \textit{International Journal of Epidemiology} (IJE) that provide detailed descriptions of HDSS sites, their data and their key results -- for example \cite{sankoh2012indepth} and \cite{kahn2012profile}.  Searching the IJE website with the phrase \textit{Health and Demographic Surveillance System Profiles} provides a quick way to find most of them.

%
%
%
%

\section{Potential Contribution of HDSS to 2030 Agenda}

The 2030 Agenda broadly calls for disaggregated population and health indicators describing national-scale populations at sub-national levels with frequent updates.  In its current form the HDSS method does all of this except describe large populations, and consequently it is worth identifying what can be learned and scaled-up from HDSS and how existing HDSS sites might contribute to the production of national-scale data.

\subsection{Cause-and-Effect}

HDSS is designed to study cause-and-effect relationships, and there will continue to be an ongoing need to conduct trials of all sorts.  HDSS sites should continue doing this, and perhaps they should be expanded and replicated to provide LMICs with additional capacity to test pharmaceuticals, vaccines and behavioral interventions locally.

\subsection{Triangulation \& Data Amalgamation}

Although many LMICs do not have traditional vital statistics and economic monitoring systems, they do have a variety of sources of data that can be combined to provide a reasonable description of the population and its health through time.  Traditional data sources include the census, a wide variety of household surveys, some economic activity surveys, administrative records, facility-based (especially health) records, HDSS and other more \textit{ad hoc} sources.  Among these HDSS data are usually the most detailed and the most accurate, but they suffer from the fact that they describe relatively small, geographically circumscribed populations.  

Combining data from multiple sources in many cases fills in temporal gaps in individual sources and covers much larger physical spaces.  This has been done in idiosyncratic ways in many places, but a set of common principles and methods has not yet emerged.  

Amalgamation of this type is possible through several general approaches: 
\begin{inparaenum}[(1)]
\item `merging' or linking data sets from different sources that describe the same or similar units of analysis and then imputing data that are missing from each source,
\item calculating indicator values and uncertainty defined in the same way from each data source and merging those into a single data set that has wider temporal and spatial extent, and 
\item applying a statistical model to a dataset of indicator values from several sources to identify and remove bias in each source and take advantage of spatial and temporal structure to `smooth' the data and thereby produce final estimates with smaller uncertainty and both spatial and temporal continuity. 
\end{inparaenum}

HDSS data offer a fourth possibility.  Because HDSS data describe the population so comprehensively, it is possible to understand relationships between indicator values that are widely collected in other sources of data, such as SES, and particular outcomes of interest.  This information can be leveraged to design \textit{very efficient, targeted} sampling designs for sample surveys that produce representative indicator values in large populations that are similar to the HDSS from which the relationship is understood.  Potentially this makes existing HDSS sites much more valuable without requiring them to change at all.  \cite{clark2015hyak,clark2018h} present a study of this idea using child mortality as the outcome and \cite{mercer2015space} apply a similar approach to HDSS data and sample survey data from the demographic and health surveys (DHS) in Tanzania. 

%
%

\subsection{Calibration of `Big Data'}

One of the most exciting things to emerge in recent years is the potential of `big data' to greatly improve the coverage, both spatial and temporal, and the content of routine data describing populations and their health.  By definition and in stark contrast to all traditional data sources, big data do not have a statistical design that dictates how they are related to the population of interest, and therefore what they \textit{mean} with respect to that population and how much variability they are expected to have.  Traditional data sources are `samples' of some kind, approaching 100\% for the census or an HDSS, and much less for household surveys.  In those cases the resulting indicator values are generalizable to the population from which the sample was drawn and uncertainty is largely related to sampling variability.  Big data of the type that could or would be used for population and health indicator production are all a byproduct of other activities that have no statistical design whatsoever, so-called `digital exhaust'.  A good example is cell phone call metadata -- the numbers from and to which a call is made along with the times when the call starts and ends and the cell towers to which the participating phones were linked at the time -- effectively a location for each phone.  This is a lot of useful information, but it pertains to people who have cell phones and use them.  Such data cannot say anything about people without cell phones, or those with phones who never use them.  Likewise, this kind of data says more about people who use their cell phones more, likely more affluent people.  It is easy to see that data of this type are biased in potentially many ways.

\textit{Perhaps the most valuable use of HDSS sites in the era of big data will be to characterize the bias in big data.}  By adding a detailed `cell phone module' to ongoing HDSS data collection it will be possible to understand cell phone ownership and usage in detail in addition to all of the other detailed information describing the HDSS study populations.  Then by combining all of this with cell phone call metadata that includes the cell phones used by the HDSS study population, it will be possible to characterize and understand the biases and omissions inherent in the cell phone call metadata, and that understanding can be used to de-bias, calibrate and adjust indicator values produced using cell phone call metadata that describe very large populations that are similar to the HDSS study population.

This general approach should work for any type of big data, social network data, satellite imagery, etc.  This is not a new concept, `ground truthing' has been practiced by geographers and mapmakers for a long time.  In this case we are ground truthing big data by calibrating them using the fact that in and HDSS we can understand both the big data and the indicators we are interested in and relate them to one another.

\subsection{Training}

Probably the most important challenge to the Data Revolution is the lack of adequately trained young people to conduct data-intensive work in LMICs where the need will be greatest.  Without a dramatic increase in data-oriented training opportunities for young people, there will simply not be enough adequately trained people to supply the data needed to monitor progress toward the SDGs.

HDSS sites are established research institutions with permanent infrastructure and sizable permanent staff with significant expertise and depth of skill and experience.  In some, but not all, cases they are already linked with universities and other national institutions engaged in either research or training or both.  Regardless, HDSS sites offer an ideal potential to contribute to training population and health professionals in all aspects of producing meaningful population and health indicators and conducting research.  HDSS sites plan studies; design and test study instruments; design, test, build and operate substantial data management systems; conduct sophisticated analysis; and publish scientific research and data/indicator datasets.  In this sense HDSS is a one-stop shop for internships and apprenticeships for trainees in a wide range of fields.  The `hands-on' nature of work at an HDSS site would add significantly to the timely production of capable new population and health research professionals.

\section{Discussion}

The strengths of HDSS include very detailed descriptions of whole populations with very frequent updates.  This often provides long time series of very accurate population and health indicators for the HDSS study population.  The primary weakness of HDSS is that the data describe \textit{only} the HDSS study population and cannot be generalized beyond that.  

With respect to the 2030 agenda, HDSS can contribute by: 
\begin{inparaenum}[(1)]
\item continuing to conduct cause-and-effect studies; 
\item contributing to data triangulation or amalgamation initiatives; \item 
characterizing the bias in and calibrating big data, and 
\item contributing more to the rapid training of data-oriented professionals, especially in the population and health fields.
\end{inparaenum}

\bibliography{HDSS-2030.bib}
\bibliographystyle{demography}

\end{document}